\documentclass[pre,aps,eqsecnum]{revtex4}
\usepackage{graphicx}
\usepackage{subfigure}
\begin{document}
%\draft
\title{Probing Spin-Polarized Currents in the Quantum Hall Regime}

\author{Thomas Herrle, Tobias Leeb, Guido Schollerer, Werner Wegscheider}

\address{Institut f\"ur Experimentelle und Angewandte Physik,
Universit\"at Regensburg, 93040 Regensburg, Germany}

\begin{abstract}

An experiment to probe spin-polarized currents in the quantum Hall regime is suggested that takes advantage of the large Zeeman-splitting in the paramagnetic diluted magnetic semiconductor zinc manganese selenide (Zn$_{1-x}$Mn$_x$Se). In the proposed experiment spin-polarized electrons are injected by ZnMnSe-contacts into a gallium arsenide (GaAs) two-dimensional electron gas (2DEG) arranged in a Hall bar geometry. We calculated the resulting Hall resistance for this experimental setup within the framework of the Landauer-B\"uttiker formalism. These calculations predict for 100\% spininjection through the ZnMnSe-contacts a Hall resistance twice as high as in the case of no spin-polarized injection of charge carriers into a 2DEG for filling factor $\nu=2$. We also investigated the influence of the equilibration of the spin-polarized electrons within the 2DEG on the Hall resistance. In addition, in our model we expect no coupling between the contact and the 2DEG for odd filling factors of the 2DEG for 100\% spininjection, because of the opposite sign of the g-factors of ZnMnSe and GaAs.       
\end{abstract}
\pacs{}
\maketitle
%\pagebreak
%\begin{multicols}{2}
%\narrowtext
%%%%%%%%%%%%%%%%%%%%%%%%%%%%%%%%%%%%%%%%%%%%%%%%%%%%%%%%%%%%%%%%%%%%
\section{Introduction}
%%%%%%%%%%%%%%%%%%%%%%%%%%%%%%%%%%%%%%%%%%%%%%%%%%%%%%%%%%%%%%%%%%%%

Since the proposal of a spintransistor by Datta and Das \cite{dat90} great efforts have been performed to experimentally realize devices using the spin as another degree of freedom in addition to the charge of an electron. The proposed spintransistor by Datta and Das consists of a field effect transistor (FET) with ferromagnetic contacts. The use of metallic ferromagnetic contacts has been questioned for achieving considerable spin-polarized currents in combination with semiconductors, because of the different conductivities of metals and semiconductors \cite{schm00}. In order to achieve considerable spin-polarized currents in the semiconductor almost 100\% intrinsic spinpolarization in the metallic ferromagnet is required, however at best 80\% spinpolarization was achieved \cite{schm00}. Therefore different other materials are examined experimentally as well as theoretically, as for example ferromagnetic semiconductors such as $\rm{GaMnAs}$ or $\rm{GaMnN}$ \cite{ohn99,kac01}. A very efficient spininjection experiment was performed by Fiederling et al. \cite{fie99} in which a spin-polarized current was injected into a light emitting diode (LED) structure through a beryllium zinc manganese selenide (BeZnMnSe)-contact. The degree of spinpolarization was determined by the measurement of the degree of the circular polarization of the emitted light. The efficiency of the spininjection was determined to be 90\%. This is due to the large Zeeman splitting of the conduction band in an applied magnetic field in this paramagnetic diluted magnetic semiconductor. We also use ZnMnSe-contacts in the consideration of our proposed experiment, where the spin-polarized current injected into a two-dimensional electron system is probed entirely electrically.

%%%%%%%%%%%%%%%%%%%%%%%%%%%%%%%%%%%%%%%%%%%%%%%%%%%%%%%%%%%%%%%%%%%%
\section{Characterization of the system}
%%%%%%%%%%%%%%%%%%%%%%%%%%%%%%%%%%%%%%%%%%%%%%%%%%%%%%%%%%%%%%%%%%%%

We consider a gallium arsenide (GaAs) two-dimensional electron system in a Hall bar geometry with ZnMnSe-contacts as schematically shown in Fig. 1. The different resistance measurements that can be performed in this Hall bar geometry can be described within the framework of the Landauer-B\"uttiker formalism provided the edge channel model can be applied, i.e. the fermi energy lies in the energy gap between two adjacent Landau or spin-split levels. 
%%%%%%%%%%%%%%%%%%%%%%%%%%%%%%%%%%%%%%%%%%%%%%%%%%%%%%%%%%%%%%%%%%
%                Figure 1:
%%%%%%%%%%%%%%%%%%%%%%%%%%%%%%%%%%%%%%%%%%%%%%%%%%%%%%%%%%%%%%%%%%
\begin{figure}
\includegraphics[width=7cm]{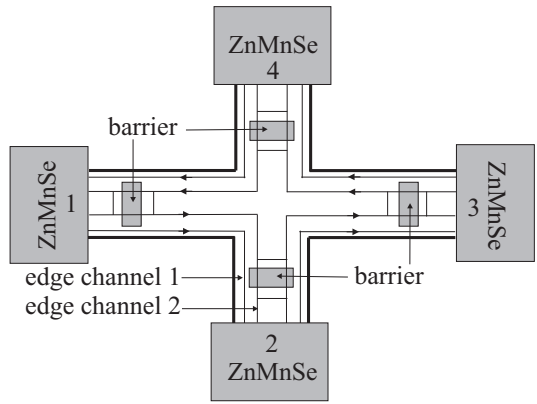} 
\caption{Hall bar geometry of a 2DEG (filling factor $\nu=2$) with ZnMnSe-contacts, which are simulated in the Landauer-B\"uttiker formalism by barriers with some transmission probability $\epsilon$.}
\label{hallbar}
\end{figure}
In this regime the current is carried by $\nu$ edge channels, where $\nu$ is the filling factor. A finite system with a certain number of contacts $J$ which are on different chemical potentials $\mu_k$ ($k=1,2,\ldots,J$) is described by the following equation \cite{bue88,bee91}:
$$ I_i=\frac{e^2}{h}\left[\left( N_i-R_i\right)V_i-\sum_{j\not=
i}T_{ji}V_j\right].\eqno(1)$$
Spin-split Landau levels are treated as separate edge channels. $I_i$ is the net current at contact $i$, $N_i$ is the number of edge channels at contact $i$, $R_i$ is the total reflection coefficient at contact $i$, $T_{ji}$ is the probability that the electrons coming from contact $i$ will reach contact $j$ and $V_i$ is the potential at contact $i$. The transmission probability $T_{ji}$ consists of the transmission probabilities $T_{ji}^{mn}$ between the separate edge channels which describe the probabilities that the electrons coming from edge channel $n$ at contact $i$ will reach the edge channel $m$ at contact $j$:
$$T_{ji}=\sum_{m,n}^{N_j,N_i}T_{ji}^{mn}.\eqno(2)$$
The total reflection coefficient $R_i$ describes what fraction of the electrons coming from contact $i$ will return to that contact and is given by:
$$R_i=\sum_{m,n}^{N_j,N_i}R_i^{mn}.\eqno(3)$$ 
Current conservation requires
$$R_i+\sum_{j\not=i}^{N_i}T_{ji}=N_i,\eqno(4) $$
$$\sum_i I_i =0. \eqno(5)$$
The four spinaligning ZnMnSe-contacts in Fig. 1 are simulated in the Landauer-B\"uttiker formalism by four barriers which are completely transparent for the majority spin-type provided by the spinaligning ZnMnSe-contact and block the minority spin-type with some probability $\epsilon$. The majority spin level can be matched to the conduction band of the two-dimensional electron system, reflected by a transmission probability of one for the majority spin-type. This is essential to be able to perform magnetotransport measurements in the quasi-equilibrium case. The transmission probability $\epsilon$ in general represents a parameter to introduce an imbalance to the edge channels of the two dimensional electron system. In our case the aim is to achieve almost 100\% spinpolarization in the ZnMnSe-contacts. The introduced barrier and the corresponding transmission probability $\epsilon$ is a measure for the deviation from the fully spin-polarized case and reflects in our model the probability of the fully spin-polarized electrons in the majority level to be thermally activated to the minority spin level. This is described in more detail in section V.
\section{Hall- and longitudinal resistance for no spin equilibration}
First we will assume that no transitions of charge carriers will occur between different edge channels, which leads to $R_i^{mn}=0$ and $T_{ji}^{mn}=0$ for all $m\not=n$. Assuming a filling factor of $\nu=2$ in the 2DEG, i.e. two edge channels are present, each edge channel carries one spin-type, namely spin up and spin down, respectively. In Fig 1. the barrier for edge channel 1 (majority spin) is completely transparent, whereas the barrier for edge channel 2 (minority spin) is only transparent to a fraction $\epsilon$. Applying equation (1) to this contact geometry leads to the following Hall resistance $R_{xy}$, when the current is applied between contacts 1 and 3 and the Hall voltage is measured between the contacts 2 and 4:
$$R_{xy}=\frac{h}{e^2}\frac{1-\epsilon+\epsilon^2}{1+\epsilon^2}\eqno(6)$$
The longitudinal resistance $R_{xx}$, when the current is applied between contacts 1 and 2 and the longitudinal voltage is measured between contacts 3 and 4, is calculated according to equation (1) as follows:
$$R_{xx}=\frac{h}{e^2}\frac{\epsilon(1-\epsilon)^2}{4(1+\epsilon^2)}.\eqno(7)$$  
The Hall resistance $R_{xy}$ and the longitudinal resistance $R_{xx}$ are shown in Fig. 2. 
%
%%%%%%%%%%%%%%%%%%%%%%%%%%%%%%%%%%%%%%%%%%%%%%%%%%%%%%%%%%%%%%%%%%
%                Figure 2:
%%%%%%%%%%%%%%%%%%%%%%%%%%%%%%%%%%%%%%%%%%%%%%%%%%%%%%%%%%%%%%%%%%
\begin{figure}
\includegraphics[width=7cm]{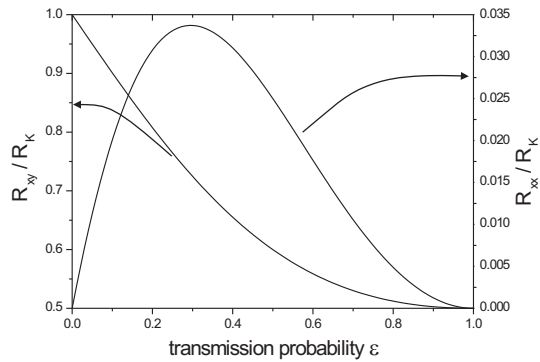} 
\caption{Hall resistance $R_{xy}$ (left axis) and longitudinal resistance $R_{xx}$ (right axis) normalized to the Klitzing resistance $R_K=\frac{h}{e^2}$ as a function of $\epsilon$ 
for filling factor $\nu=2$, neglecting spin equilibration. The Hall resistance $R_{xy}$ decays with increasing transmission probability $\epsilon$ from $R_K$ to $\frac{R_K}{2}$, reflecting the decreasing imbalance of the two edge channels for $\epsilon\rightarrow 1$. The non-vanishing longitudinal resistance $R_{xx}$ can be explained by backscattering of electrons (see text).}
\label{hallbar}
\end{figure}
If the ZnMnSe-contacts were completely transparent for both spin-types ($\epsilon=1$), the Hall resistance would be $R_{xy}=\frac{R_K}{2}$, with $R_K=\frac{h}{e^2}$ and the longitudinal resistance would be $R_{xx}=0$. In the case of complete spin-polarized injection ($\epsilon=0$), the Hall resistance and longitudinal resistance turn out to be $R_{xy}=R_K$ and $R_{xx}=0$, respectively. For $\epsilon\not=\{0,1\}$ the Hall resistance $R_{xy}$ decays from $R_{xy}=R_K$ to $R_{xy}=\frac{R_K}{2}$, reflecting the fact that the imbalance of the two edge channels is decreased as $\epsilon\rightarrow 1$. The non-vanishing longitudinal resistance $R_{xx}$ for values of $\epsilon\not=\{0,1\}$ can be explained by a non-vanishing backscattering probability of the electrons injected in edge channel 2 at the barriers, since edge channels at opposite sides of the sample carry the current in opposite directions (see Fig. 1). If the transmission probability $\epsilon=1$, the barriers are all completely transparent and no backscattering can occur. If $\epsilon=0$, the barriers completely block the electrons coming from contact 1 and they are backscattered into contact 1 without reaching the voltage probing contacts 3 and 4. Consequently, the measured longitudinal resistance is $R_{xx}=0$. For $\epsilon\not=\{0,1\}$ a fraction of electrons being injected through contact 1 into edge channel 2 can be backscattered and can reach the voltage probing contact 4. A non-vanishing longitudinal resistance has already been reported in gated Hall bar structures \cite{was88, kom89, rya93} and explained by a backscattering of electrons.
\section{Influence of Spin Equilibration on Hall- and Longitudinal Resistance}
The influence of electron scattering between two neighboring edge channels, i.e. an equilibration of electrons between differently populated spin-split levels can be treated within the Landauer-B\"uttiker formalism by introducing a probability $P(l)$ that an electron will remain in the same edge channel on the way between two neighboring contacts, separated by a distance $l$. To obtain this probability one assumes that two edge channels are populated. In our case namely a spin-up and a spin-down edge channel. A constant number of inter edge channel scattering events is assumed, $\frac{1}{l_{eq}}$, where $l_{eq}$ is the equilibration length of the electrons, which is the distance between two edge channel electron scattering processes. Now rate equations for the population of the edge channels can be written down. If $P_\downarrow$ is the probability to find an electron in the spin down edge channel and $P_\uparrow$ the corresponding probability for the spin up edge channel one gets:
$$\frac{\mathrm{d}P_\downarrow}{\mathrm{d}l}=-\frac{1}{l_{eq}}P_\downarrow +\frac{1}{l_{eq}}P_\uparrow , \eqno(8)$$
$$\frac{\mathrm{d}P_\uparrow}{\mathrm{d}l}=\frac{1}{l_{eq}}P_\downarrow -\frac{1}{l_{eq}}P_\uparrow .\eqno(9)$$
With the boundary condition at $l=0$ of $P_\downarrow=1$ and $P_\uparrow=0$ one obtains the following solutions of the differential equations (8) and (9):
$$P_\downarrow=\frac{1}{2}+\frac{1}{2}\exp\left(-\frac{2l}{l_{eq}}\right) , \eqno(10)$$
$$P_\uparrow=\frac{1}{2}-\frac{1}{2}\exp\left(-\frac{2l}{l_{eq}}\right) . \eqno(11)$$ 
Similar to Ref.~\onlinecite{mue92} the probability of an electron remaining in one edge channel is therefore given by $P(l)=P_\downarrow$.
The probability that an electron will be scattered on its way between two contacts into a neighboring edge channel is then simply given by $1-P(l)=P_\uparrow$. For a contact geometry according to Fig. 1 and a filling factor of $\nu=2$ one can calculate the resulting Hall resistance with the help of equation (1)
$$R_{xy}=\frac{h}{e^2}\frac{P(l)^2(\epsilon-1)^2+\epsilon}{2P(l)^2(\epsilon-1)^2-2P(l)(\epsilon-1)^2+\epsilon^2+1} \eqno(12)$$
and the resulting longitudinal resistance  
$$R_{xx}=-\frac{h}{e^2}\frac{1}{4}\frac{2P(l)^2(\epsilon-1)^3-P(l)(\epsilon^3-5\epsilon^2+7\epsilon-3)-(\epsilon-1)^2}{[2P(l)^2(\epsilon-1)^2-2P(l)(\epsilon-1)^2+\epsilon^2+1][P(l)(\epsilon-1)-\epsilon]}. \eqno(13)$$
In the experiment performed by M\"uller et al. \cite{mue92}, where the two spin-polarized edge channels of the lowest Landau level ($\nu=2$) in a gallium arsenide/aluminum gallium arsenide (GaAs/Al$_x$Ga$_{1-x}$As) 2DEG are selectively populated by applying a negative gate bias to Schottky gates on top of a Hall bar geometry, typical spin-flip equilibration lengths $l_{eq}$ between 100\,$\mu$m and 1\rm{\,mm} were found at a temperature of 100\rm{\,mK}, which decrease significantly at temperatures above 250\rm{\,mK}. These values for the equilibration lengths are also supported by theoretical findings reported in Ref.~\onlinecite{kha92}, where the spin-flip process is mediated by spin-orbit interaction. The Hall resistance $R_{xy}$ and the longitudinal resistance $R_{xx}$ as a function of the transmission probability $\epsilon$ of the ZnMnSe-contacts and their distance $l$ for an equilibration length of $l_{eq}=200\,\mu$m are shown in Fig. 3a) and Fig. 3b), respectively. 
%%%%%%%%%%%%%%%%%%%%%%%%%%%%%%%%%%%%%%%%%%%%%%%%%%%%%%%%%%%%%%%%%%
%                Figure 3:
%%%%%%%%%%%%%%%%%%%%%%%%%%%%%%%%%%%%%%%%%%%%%%%%%%%%%%%%%%%%%%%%%%
\begin{figure}[!hbt]
\centering
\subfigure[ Hall resistance]{\label{hall_e_x_n2}
\includegraphics[width=7cm]{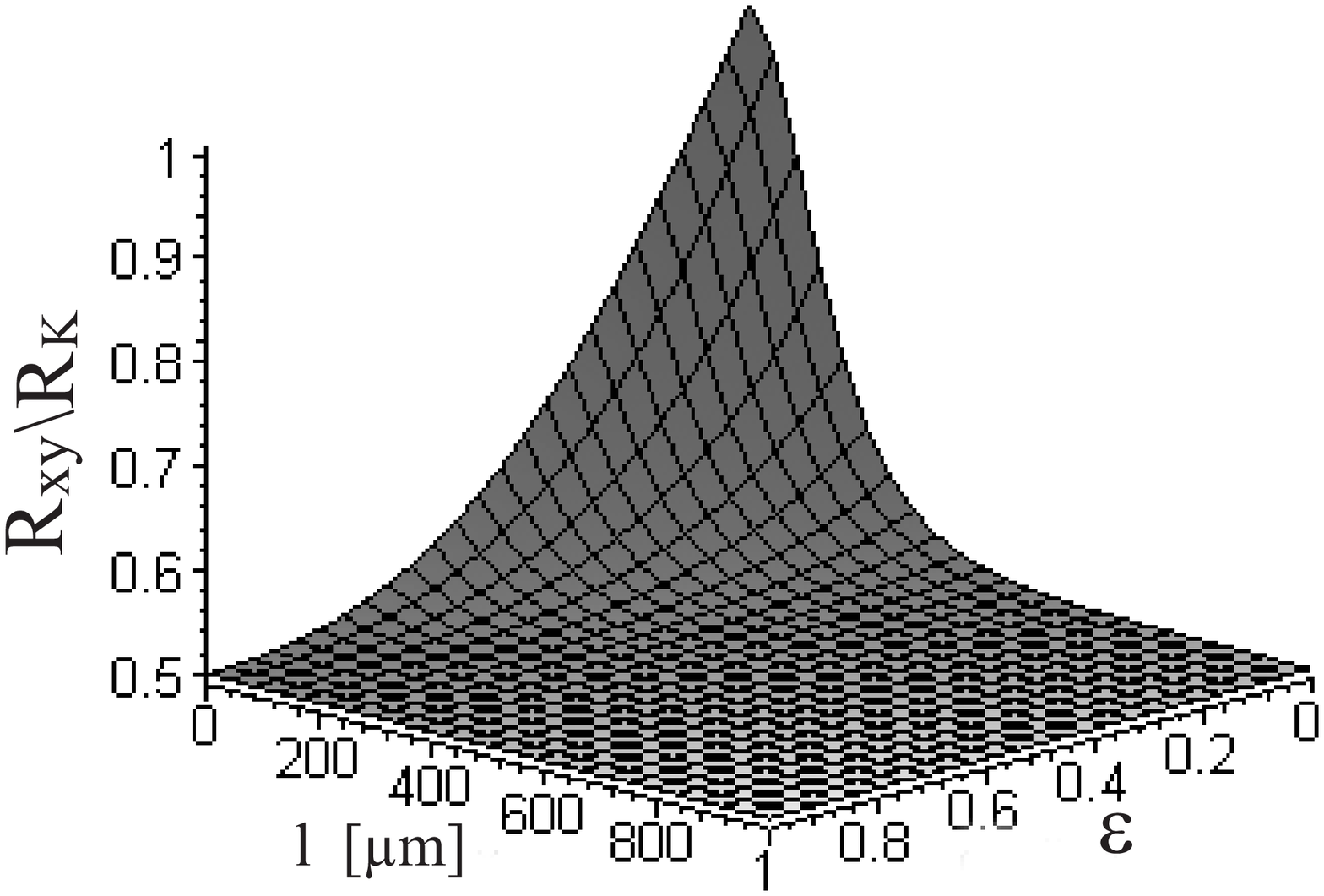}}
\subfigure[ Longitudinal resistance]{\label{laengs_e_x_n2}
\includegraphics[width=7cm]{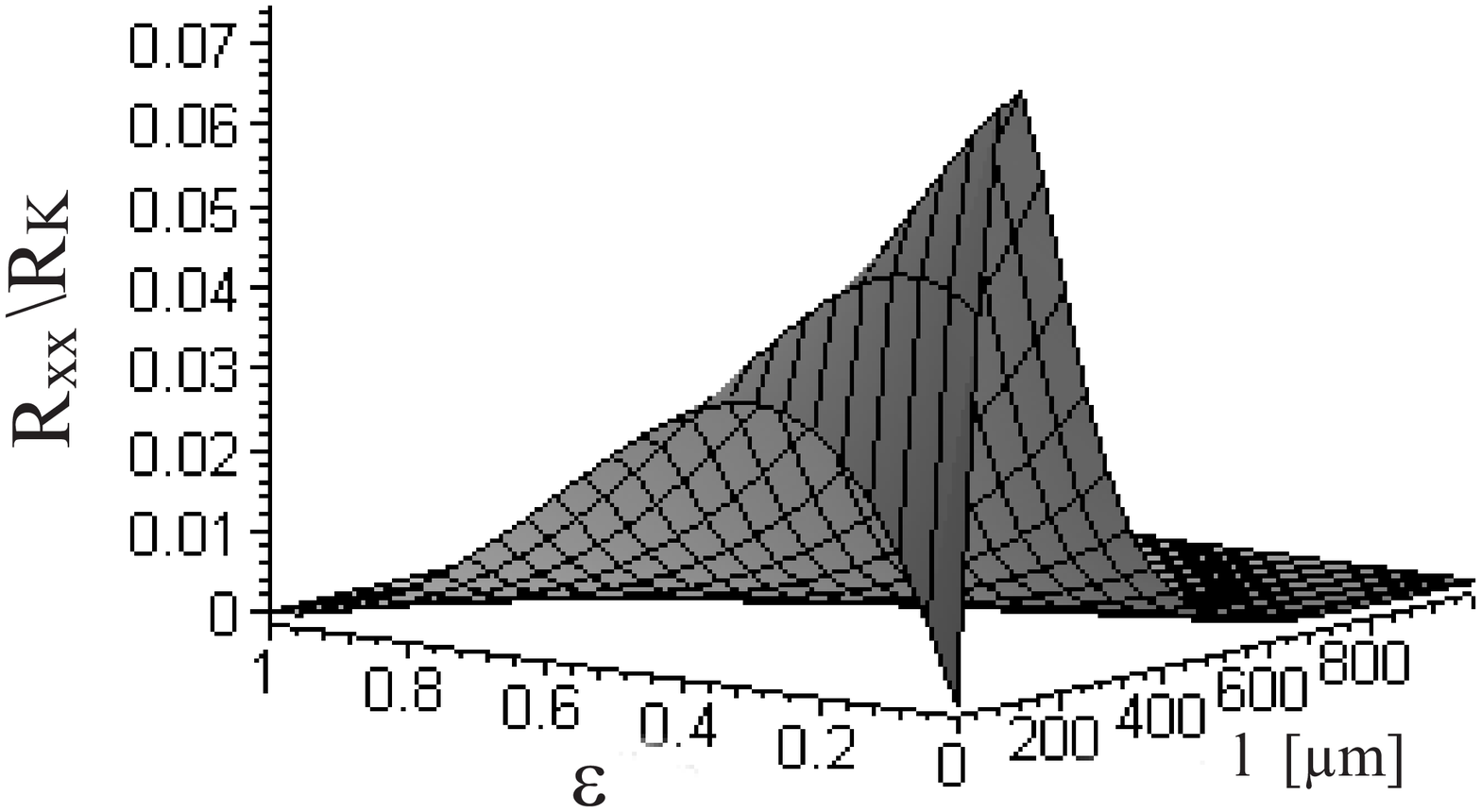}}
\caption{a) Normalized Hall resistance $R_{xy}$ and b) normalized longitudinal resistance $R_{xx}$ as a function of $\epsilon$ and the distance $l$ between the ZnMnSe-contacts of a 2DEG with filling factor $\nu=2$ for an equilibration length of $l_{eq}=200\,\mu\mathrm{m}$. The Hall resistance $R_{xy}$ decays exponentially with increasing distance of the ZnMnSe-contacts, reflecting the equilibration of the two edge channels. The longitudinal resistance $R_{xx}$ vanishes for $l=0$ and approaches zero for infinite contact distances $l$. The non-vanishing longitudinal resistance for a finite contact distance $l$ can be explained by the backscattering of electrons (see text).}
\end{figure} 
%%%%%%%%%%%%%%%%%%%%%%%%%%%%%%%%
For complete spininjection ($\epsilon=0$) the Hall resistance decays exponentially with the distance $l$ of the ZnMnSe-contacts from $R_K$ to $\frac{R_K}{2}$, reflecting the equilibration of the two spin-split edge channels. The non-vanishing longitudinal resistance for complete spininjection ($\epsilon=0$) with distance $l$ can be explained by the scattering of electrons between neighboring edge channels. Electrons can be scattered from edge channel 1 into edge channel 2 and can therefore be backscattered, leading to a non-vanishing longitudinal resistance. For longer distances $l$ between the contacts complete equilibration of electrons is reached and the longitudinal resistance again vanishes.   
\section{Magnetic Field Dependent Hall Resistance}
In order to describe the magnetic field dependence of the Hall resistance of the system depicted in Fig. 1, equation (12) has to be extended to different filling factors $\nu$. For this purpose the transmission- and reflection probabilities of equations (2) and (3), respectively, have to be re-calculated. The transmission probability $\epsilon$ of the ZnMnSe-contacts does depend on the magnetic field. The aim is to achieve almost 100\% spinpolarization in the ZnMnSe-contacts, i.e. the minority spin level is above the Fermi level in the contacts. This can be achieved by suitable doping of the ZnMnSe-contacts. In this doping regime the majority and minority spin levels can be treated as a two level system, separated by approximately the Zeeman energy $\Delta E_z=g_{eff}\mu_B B$. $g_{\mathrm{eff}}$ is the effective g-factor which is described in Refs.~\onlinecite{fur-a88} and \onlinecite{fur-b88}:
$$g_{\mathrm{eff}}=g^*+\frac{N_0\alpha x}{\mu_BB}S_{\mathrm{eff}}(x)\mathrm{B}_\frac{5}{2}(y_1)\approx 2+\frac{N_0\alpha x}{\mu_BB}S_{\mathrm{eff}}(x)\mathrm{B}_\frac{5}{2}(y_1).\eqno(14)$$ 
$g^*\approx 2$, $y_1=\frac{5}{2}\frac{g_{Mn}\mu_B}{k_B\left(T+T_{\mathrm{eff}}(x)\right)}$ and $g_{Mn}=2$. The manganese content of the Zn$_{1-x}$Mn$_x$Se-contacts is $x$ and $N_0\alpha\approx0,26\mathrm{\,eV}$. $S_{\mathrm{eff}}$ and $T_{\mathrm{eff}}$ are phenomenological parameters, which depend on the Mn-content $x$. $S_{\mathrm{eff}}$ accounts for the fact that a fraction of the Mn$^{2+}$-ions in the ZnMnSe lattice will build pairs or even larger clusters which interact antiferromagnetically and thus decrease the effective magnetic moment. This fraction of Mn$^{2+}$-ions increases with increasing Mn-content in the ZnMnSe-contacts. The remaining ions which are not in pairs or clusters are aligned to an external magnetic field according to a Brillouin function $\mathrm{B}_\frac{5}{2}(y_1)$ with a temperature $T+T_{\mathrm{eff}}$ in the exponent which is higher than the actual sample temperature $T$. The reason for this is a long range antiferromagnetic interaction between these ions which also increases with increasing Mn concentration in the ZnMnSe-contacts \cite{gaj79}. Considering the two level system of the majority and minority spin levels, where the minority spin level lies above the Fermi level, the minority spin level can be thermally activated. Using a Boltzman approximation, where the ratio of the minority and the majority spin population is given by $\frac{n_\uparrow}{n_\downarrow}=\exp\left(-\frac{\Delta E_z}{k_BT}\right)$, the transmission probability $\epsilon$, which reflects the deviation from the full spinpolarization of the ZnMnSe-contacts, can be written as 
$$\epsilon=1-\frac{n_\downarrow-n_\uparrow}{n_\downarrow+n_\uparrow}=\frac{2}{1+\exp\left(\frac{\Delta E_z}{k_BT}\right)}.\eqno(15)$$
The transmission probability $\epsilon$ of the minority spin through the ZnMnSe-contact drops down to $4,2\cdot10^{-350}$ at a magnetic field of $B=2\mathrm{\,T}$ at 100\rm{\,mK} for a manganese content of $x=0,032$. This high degree of spinpolarization even at small magnetic fields is consistent with the experiments of Ref.~\onlinecite{fie99} and \onlinecite{schm01} and may be explained by conduction in an impurity band, which is split by the Zeeman energy into a two level system of spin up and spin down electrons in an applied magnetic field as discussed above. That is the reason for the high degree of spinpolarization in ZnMnSe already at small magnetic fields. Since it is possible to achieve full spinpolarization at a magnetic field of $B=2\mathrm{\,T}$ for suitable doping of the ZnMnSe-contacts and $\epsilon$ is essentially zero for $B>2\mathrm{\,T}$ we assume $\epsilon=0$ for these magnetic fields in the following simulation, i.e. 100\% spinpolarization of the ZnMnSe-contacts. Spin-flip scattering at the interface between ZnMnSe and the 2DEG and also on the way between the ZnMnSe contact and the 2DEG is neglected. The latter is justified, since long spin decoherence lengths have been found in GaAs \cite{kik99}. When simulating the magnetic field dependence of the Hall resistance $R_{xy}$ one has to compute the filling factor $\nu$ of the 2DEG for the applied magnetic field $B$ using $\nu=\frac{n_s h}{eB}$, where $n_s$ is the charge carrier density in the 2DEG. The knowledge about the filling factor, i.e. knowing the number of edge channels, enables us to compute the transmission and reflection probabilities for different magnetic fields according to equations (2) and (3). From equation (1) the Hall resistance $R_{xy}$ can now be deduced. In this model only scattering between neighboring spin-split edge channels is considered, neglecting scattering between different Landau levels. This is justified since the Zeeman-splitting in the GaAs is much smaller than the Landau splitting and thus the overlap of the wave functions of the spin-split levels is larger than that of the Landau levels \cite{mue92}. We also neglect possible spin-flip scattering events at the interface between the ZnMnSe-contacts and the 2DEG. In our simulation of the magnetic field dependence of the Hall resistance for a contact geometry according to Fig. 1 we use an equilibration length of $l_{eq}=200\,\mu\mathrm{m}$ which is a typical value reported in Ref.~\onlinecite{mue92}. In our model we assume this equilibration length to be the same for all filling factors, which is certainly not true since the overlap of the wave functions of spin-split levels increases with decreasing magnetic field, i.e. with increasing filling factors. This is due to the reduction of the Zeeman-splitting in the GaAs with decreasing magnetic field. We use this assumption, which is quite good for small filling factors, because of the lack of the knowledge of the equilibration length for filling factors higher than two. The temperature of the sample is considered to be $T=100\mathrm{\,mK}$, the charge carrier density is assumed to be $n_s=7\cdot10^{15}\mathrm{\,m}^{-2}$ and the Mn-concentration of the ZnMnSe-contacts is set to $x=0,032$. The result is shown in Fig. 4, where the Hall resistance $R_{xy}$ is displayed versus the magnetic field for different distances $l$ between the ZnMnSe-contacts. In the Landauer-B\"uttiker formalism the Hall resistance arises from counting the current carrying spin-split Landau levels, i.e. the current carrying spin-split edge channels at the Fermi level in the 2DEG. Each of these edge channels contributes $\frac{h}{e^2}$ to the Hall resistance. 
%%%%%%%%%%%%%%%%%%%%%%%%%%%%%%%%%%%%%%%%%%%%%%%%%%%%%%%%%%%%%%%%%%%%%%%%%%%%%%%%%%%%%%%%%%%%%
% Figure 4:
%%%%%%%%%%%%%%%%%%%%%%%%%%%%%%%%%%%%%%%%%%%%%%%%%%%%%%%%%%%%%%%%%%%%%%%%%%%%
\begin{figure}[!hbt]
\center
\includegraphics[width=7cm]{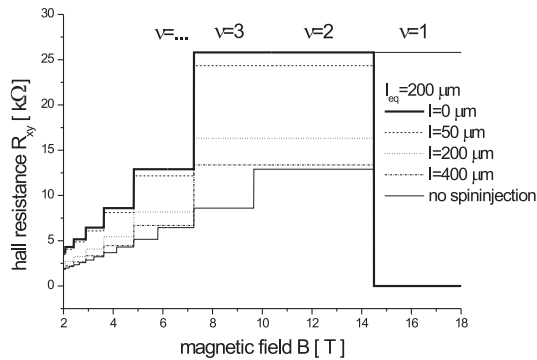}
\caption{Hall resistance $R_{xy}$ as a function of the applied magnetic field $B$ for different distances $l$ between the ZnMnSe-contacts for an equilibration length of $l_{eq}=200\mathrm{\,\mu m}$. With increasing distance $l$ of the ZnMnSe-contacts the Hall resistance $R_{xy}$ decays exponentially to approach the  value of the Hall resistance for not spin-polarized contacts (thin solid line). The reason that the Hall resistance for odd filling factors does not decay to the value for the case of not spin-polarized contacts is the opposite sign of the g-factor in ZnMnSe and GaAs and the fact that the presented model does not include inter Landau level scattering (see text).}
\label{hallsim4kont}
\end{figure}
%%%%%%%%%%%%%%%%%
The case of no spininjection is also displayed in Fig. 4 for comparison (thin solid line). Without equilibration of electrons between the spin-split levels ($l=0\,\mu\mathrm{m}$, bold solid line), the Hall resistance would be higher than in the case of no spininjection. For a filling factor of $\nu=2$ the Hall resistance for no equilibration of spins ($R_{xy}=R_K$) is twice as high as in the case of no spininjection ($R_{xy}=\frac{R_K}{2}$). With increasing distance $l$ between the ZnMnSe-contacts the Hall resistance decreases and exponentially approaches the value of the Hall resistance for no spininjection in the case of even filling factors $\nu$. For complete spininjection for $\nu=2$ there is just one current carrying edge channel at the Fermi level, namely the channel for the majority spin. With increasing distance $l$ of the ZnMnSe-contacts, spin-flip processes lead to a population of the second edge channel, i.e. the edge channel corresponding to the minority spin-type. Therefore one ends up with two current carrying edge channels at the Fermi level in the 2DEG. For a filling factor of $\nu=1$ the electrons in the GaAs have an opposite spin compared to the majority electrons in the ZnMnSe-contact, because of the opposite sign of the g-factor of GaAs and ZnMnSe. Therefore we expect no current injection from the ZnMnSe-contact into the $\nu=1$ level of the 2DEG and no Hall voltage is expected to be measured for 100\% spininjection. This effect is destroyed for not 100\% spininjection ($\epsilon\not=0$). For odd filling factors $\nu$ the Hall resistance in our model does not decay to the value of the case of no spininjection. This is also due to the opposite sign of the g-factor in ZnMnSe and GaAs. In the case of filling factor $\nu=3$, for example, two Landau levels have an opposite spin compared to the majority electrons in the ZnMnSe-contact. There is effectively just one level in the 2DEG for the majority spin in the ZnMnSe-contact and a second, spin-split level to which electrons can be equilibrated, when neglecting scattering between different Landau levels. This means that there is just one current carrying channel for no equilibration and two of these channels, which contribute to ballistic transport, with different population for finite equilibration between spin-split levels. This is the reason why the Hall plateau corresponding to filling factor $\nu=3$ is at the same level as the Hall plateau for $\nu=2$. If scattering between Landau levels was included, the Hall resistance would again decay to the value of the case of no spininjection, however, according to a probably longer equilibration length for this process, on a different length scale. A deviation from the case of full spininjection ($\epsilon\not=0$) will cause the Hall plateaus corresponding to different Landau levels of opposite spin, for example $\nu=3$ and $\nu=2$, not to align perfectly. The deviation will depend on the value of $\epsilon$.\\
An important point to note here is that spin accumulation effects and band bending are not considered in the presented model. These effects lead to a filling of the minority spin level in the ZnMnSe-contacts and thus to an increase in the transmission probability $\epsilon$ for the minority spin-type. These effects are considered to be more pronounced at higher bias voltages. In our model we are in a regime with low bias, because a quasi equilibrium quantum Hall measurement is assumed. Nevertheless a small increase in $\epsilon$ might lead to a coupling of the minority spin-type in the ZnMnSe-contacts to the $\nu=1$ level of the 2DEG and a Hall voltage of $R_{xy}=R_K$ becomes measurable. Furthermore the Hall plateaus corresponding to different Landau levels of opposite spin will not perfectly align anymore for $\epsilon\not=0$ (see above). However, the filling of the minority spin level by spin accumulation and band bending can be prevented or kept small, by keeping the intrinsic spinpolarization of the ZnMnSe-contacts high. This can be done by introducing a sufficiently high Mn-concentration to the ZnMnSe-contacts.               
%%%%%%%%%%%%%%%%%%%%%%%%%%%%%%%%%%%%%%%%%%%%%%%%%%%%%%%%%%%%%%%%%%%%
\section{Conclusion}
%%%%%%%%%%%%%%%%%%%%%%%%%%%%%%%%%%%%%%%%%%%%%%%%%%%%%%%%%%%%%%%%%
In summary, we have proposed an experiment to probe spin-polarized currents in the quantum Hall regime of ZnMnSe-contacts coupled to a 2DEG, where we expect a Hall resistance twice as high as for conventional non spinaligning contacts for a filling factor of $\nu=2$. ZnMnSe-contacts are chosen, because of their large Zeeman-splitting and therefore large internal spinpolarization in an applied magnetic field. Furthermore the influence of the equilibration length between spin-split levels in the 2DEG was discussed, with the conclusion that the distance of the contacts has to be considerably smaller than a minimal equilibration length of $l_{eq}=100\,\mu\mathrm{m}$ at temperatures below $T=250\mathrm{\,mK}$ in order to observe the proposed effect. In our model coupling of the majority spin in the ZnMnSe-contact to odd filling factors in the 2DEG is strongly suppressed for full spininjection, because of the opposite sign of the g-factors of ZnMnSe and GaAs. The described model can also be applied to other spin-polarizing contacts, for example GaMnAs. Since this ferromagnetic semiconductor is p-type, one has to inject spin-polarized holes into a two dimensional hole gas. In order to be applicable, our model has to be adjusted for equilibration lengths in a two dimensional hole gas.   
%%%%%%%%%%%%%%%%%%%%%%%%%%%%%%%%%%%%%%%%%%%%%%%%%%%%%%%%%%%%%%%%%%%%
\section{Acknowledgment}
%%%%%%%%%%%%%%%%%%%%%%%%%%%%%%%%%%%%%%%%%%%%%%%%%%%%%%%%%%%%%%%%%

It is a pleasure to thank M.~Langenbuch and M.~Zoelfl for helpful discussions. 
Financial support from the Deutsche Forschungsgemeinschaft
SFB 348 and FOR 370 and from BMBF project 13N8282
is gratefully acknowledged.

\medskip

%%%%%%%%%%%%%%%%%%%%%%%%%%%%%%%%%%%%%%%%%%%%%%%%%%%%%%%%%%%%%%%%%%%

\end{document}